\documentstyle[12pt]{article}
\let\myfrac=\frac
\def\myfrac#1#2{\mbox{\footnotesize$\frac{\displaystyle #1}
             {\displaystyle #2}$}}

\def\numb(#1){\setcounter{equation}{#1}}

\begin{document}
\parindent 0pt

{\bf\centering   ON A POSSIBILITY OF SCALAR\\
                  GRAVITATIONAL WAVE DETECTION\\
            FROM THE BINARY PULSAR PSR $\bf 1913 \bf+16$\\}

\vskip 10mm  plus 5mm
{\bf\centering Yu.V.Baryshev\\}

{\footnotesize\it\centering
  Astronomical Institute of the Saint-Petersburg State University,\\
  Stary Petergof,  St.Petersburg, 198904, Russia;\\
  Scientific-Educational Union ``Earth \& Universe",\\
   St.Petersburg, Russia\\
              e-mail: yuba@astro.spbu.ru\\}
\vskip 10mm   plus 5mm

{\bf\hfil ABSTRACT}
\vskip 1mm

\hfil\parbox{30pc}
{\footnotesize\parindent  5mm It is shown that  detecting or
       setting an upper limit
       on the scalar gravitational radiation is a good experimental
       test
       of relativistic gravity theories. The relativistic
       tensor-field theory of gravitation is revised and it is
       demonstrated that the scalar monopole
              gravitational radiation must be added to the usual
       quadrupole radiation. In the case of the binary pulsar
       PSR $1913+16$ it is predicted the existence  0.735\%
       excess of the gravitational radiation due to the
       scalar gravitational waves.}
\vskip 10mm   plus 5mm

\parindent  12mm

\noindent
{\bf
1. Scalar Gravitational Radiation as a Test of Gravity Theories}
\vskip 5mm

\noindent
{\it
1.1. Geometry of Spacetime or Quantum Field in Flat Spacetime}
\vskip 5mm

       Gravitational  wave  experiments
give us a new possibility in choosing  among
alternative gravity theories. Here we revise
nonmetric tensor-field theory~(TFT) of gravitation, i.e.  relativistic
theory  of  symmetrical  second  rank  field
$
\psi^{ik}
$    in  flat
Minkowski spacetime
$
\eta^{ik}
$
and compare its
gravitational wave
predictions
with general relativity~(GR) ones.

         It~is  easy  to  show  that
       widespread  opinion  on  full coincidence of~TFT and GR
(see for example$^1$) based on
the assumption of
the uniqueness of the energy-momen\-tum  tensor~(EMT) of the
gravitational field, which was used  for  the  iteration procedure.
However,  in  the  framework  of
Lagrangian  formalism  of
relativistic field theory, EMT of any field is not  defined  uniquely:
$
T^{ik} \Rightarrow T^{ik} + \Phi^{ikl}_{\phantom{ikl},l}
$
 for
$
\Phi^{ikl} = - \Phi^{ilk}
$ (see~for example$^{2,\,3}$), and one needs
additional physical restrictions to choose final form of the EMT
(for example   such conditions as energy positiveness, tracelessness,
symmetry). It~is apparently that different  EMTs  will lead to
different nonlinear theories of gravitation.

To  illustrate the above-said
discussion  let  us consider nonlinear generalization of Poisson's
equation  for  the case  of  distributed source  describing  negative
and  positive energy  density   of   the gravitational field
\eject
$$
\Delta\varphi =
  - (\nabla \varphi)^2/c^2,
\eqno(1a)
$$
and
$$
\Delta\varphi =
   + (\nabla \varphi)^2/c^2.
\eqno(1b)
$$
The solution of Eq.1a is
$$
\varphi = c^2 \ln
       \left(1 - \frac{GM}{c^2r}\right), \quad
\frac{d\varphi}{dr} = \frac{GM}
{r^2\left(1-{GM}/{c^2r} \right)};
\eqno(2)
$$
whereas the solution of Eq.1b is
$$
\varphi = - c^2 \ln
       \left(1 + \frac{GM}{c^2r}\right), \quad
\frac{d\varphi}{dr} = \frac{GM}
{r^2\left(1+{GM}/{c^2r} \right)}.
\eqno(3)
$$
>From Eq.2 and Eq.3 we see two possible ways for the construction  of
the nonlinear~TFT. The first one, based on the negative energy
density of the gravitational field, leads to the infinite gravity
force  at the finite distance $ R_{g} = GM /c^2 $.  The second way,
based on positive one, gives~TFT without singularity.

As for~GR, the energy density of the gravitational field is  a
poorly defined concept as a consequence of geometrical  interpretation
of gravity (there is no tensor characteristic of  the  gravitational
field~EMT). For example, Landau-Lifshiz pseudotensor gives
$
t^{00} =  - 7(\nabla \varphi_{_{N}})^2 / 8\pi G
$
but Grischuk-Petrov-Popova one gives
$
t^{00} = - 11(\nabla \varphi_{_{N}})^2 / 8\pi G
$
for the spherically symmetric
static
(SSS)
weak field in harmonic
coordinates.

       Within the scope of TFT there is a real tensor quantity  for
the energy density of the gravitational field. For example  canonical
EMT gives
$ T^{00} = +(\nabla \varphi_{_{N}})^2 /8\pi G $
for SSS weak
field  and corresponds quantum description of the gravitational field
as aggregate  of gravitons  in flat spacetime. Gravitons  are
massless particles, i.e.  some kind of matter  in  spacetime,
 which
carry-over positive energy  and momentum in the spacetime. Besides the
sum of two tensors
$ \psi^{ik} + \eta^{ik}  = g^{ik} $ is not a
metric tensor
because the covariant components
$ \psi_{ik} + \eta_{ik}  = g_{ik} $
of this tensor provide the mixed components $ \psi_{k}^{i} +
\eta_{k}^{i}  = g_{k}^{i} \not\equiv \delta^{i}_{k} $ and the trace
$
g^{ik}  g_{ik} = 4 + 2\psi + O(\psi^{2}_{ik}) \not\equiv  4.
$ It
means that TFT  is  a scalar-tensor theory, but not pure tensor one,
and hence
includes
spin~2
and
spin~0 gravivtons.
\vskip 10mm

\noindent
{\it
   1.2. Post-Newtonian Tensor-Field Theory of Gravitation}
\vskip 5mm

       Let us consider the relativistic symmetric  tensor  field
$
\psi^{ik}
$
in Minkowski's spacetime
$
\eta^{ik}
$.
As so as   really  observed
gravitational fields are the weak ones ($|\varphi| \ll c^2$), it is
naturally to
begin the
construction of TFT for the weak field case.
In this case we
have a very closed  analogy with  the  electromagnetic  field  and
can use the standard
Lagrangian formalism of  the  relativistic
field theory (bellow  we  utilize notations of the
text-book$^2$).

We begin with the action integral in the form
$$
S = S_{(g)} +
S_{(int)} + S_{(p)} = \frac{1}{c} \int(\Lambda_{(g)} +
\Lambda_{(int)} + \Lambda_{(p)}) d\Omega \eqno(4)
$$
where~($g$),~($int$)~and~($p$) indicate gravitational field, interaction
and particles parts of actions and Lagrangians.
The Lagrangians  are
given by the following expressions
\begin{eqnarray}
 \Lambda_{(g)}\phantom{(i} &=& -
 \frac{1}{16 \pi G} \left[ 2\psi_{nm}^{\phantom{nm},n}
  \psi^{lm}_{\phantom{lm},l} - \psi_{lm,n}\psi^{lm,n} -
 2\psi_{ln}^{\phantom{ln},l} \psi^{,n} + \psi_{,l}\psi^{,l}\right],
\numb(5)
                                    \\
  \Lambda_{(int)} & = & - \frac{1}{c^2} \psi_{lm} T_{(p)}^{lm},
\numb(6)
                                     \\
  \Lambda_{(p)}\phantom{(i} & = & - \eta_{ik}T_{(p)}^{ik}.
\numb(7)
\end{eqnarray}

It has been shown by Kalman$^4$ and Thirring$^5$, that  the  total~EMT
of  the  system contains three parts, which correspond to ones  of
the  action integral~(4),
$$
T_{(\Sigma)}^{ik} = T_{(p)}^{ik} + T_{(int)}^{ik} +
   T_{(g)}^{ik}
\eqno(8)
$$
where the canonical~EMT of the gravitational field for Lagrangian~(5)
has the form
$$
T_{(g)}^{ik} = \frac{1}{8 \pi G}
 \left\{
      (\psi^{lm,i}\psi_{lm}^{\phantom{lm},k} -
 \frac{1}{2}
             \eta^{ik} \psi_{lm,n} \psi^{lm,n}) -
 \frac{1}{2}
     (\psi^{,i}\psi^{,k} - \frac{1}{2}
                 \eta^{ik}\psi_{,l}\psi^{,l})\right\}
\eqno(9)
$$
the interaction EMT is
$$
T^{ik}_{(int)} = \frac{2}{c^2}T_{(p)l}^{\phantom{(p)}i}
\psi^{lk} -\frac{1}{c^2}T^{ik}_{(p)}\psi_{lm}u^l u^m,
\eqno(10)
$$
the point particles~EMT is
$$
T_{(p)}^{ik} = \sum_{a} m_a c^2
                 \delta({\bf r}-{\bf r}_a)
     \{1 - \frac{v_a^2}{c^2}\}^{1/2}
                         u_a^i u_a^k  .
\eqno(11)
$$
Therefore the nonlinear~TFT must include the interaction Lagrangian in
the form
$$
\Lambda_{(int)} = -\frac{1}{c^2} \psi_{lm}T^{lm}_{(\Sigma)} .
\eqno(12)
$$
But the weak field condition allows us to use linear approximation
as the first step and then  to  make  nonlinear  corrections
(Post-Newtonian~TFT).

      The variation of the  gravitational  potentials  in  the  action
integral~(4), where for fixed sources in the~PN~approximation  we  can
use the interaction Lagrangian~(12), yield the PN~field  equations  in
the form
$$
-\psi_{\phantom{ik,}l}^{ik,l} + \psi_{\phantom{ik,}l}^{il,k}
    + \psi_{\phantom{ik,}l}^{kl,i} - \psi^{,ik}
       -\eta^{ik} \psi^{lm}_{\phantom{lm},lm} +
                 \eta^{ik}  \psi_{\phantom{,}l}^{,l} =
\frac{8 \pi G}{c^2} \, T_{(\Sigma)}^{ik}
\eqno(13)
$$

       Eq.13 automatically requires conservation of the total~EMT
and leads to
the particles motion equations  in the form
$
T^{ik}_{(\Sigma),k} = 0
$.

       The field equations are invariant (for fixed sources) under the
gauge transformation
$
\psi^{ik} \Rightarrow \psi^{ik} + \theta^{i,k} + \theta^{k,i}
$
and  one  can
achieve Hilbert gauge in the form:
$
\psi^{ik}_{\phantom{ik},k} = \frac{1}{2} \psi^{,i}
$.
In  this  case
the  field Eq.13 become
$$
\Box\psi^{ik} = \frac{8\pi G}{c^2}
     \left[ T^{ik}_{(\Sigma)}
 -\frac{1}{2} \eta^{ik}T_{(\Sigma)}\right] ;
\eqno(14)
$$

The  very  important  feature  of   Eq.13  is  the  multi
component structure of the gravitational  field  description.  Initial
symmetric tensor $\psi^{ik}$    simultaneously  describes  a
mixture of four particles
$$
\{\psi^{ik}\} = \{2\} \oplus \{1\}  \oplus
  \{0\} \oplus  \{0^{'}\}
\eqno(15)
$$
with spins equal~2,~1,~0  and  $0^{'}$.  It  corresponds  to~10
independent components of the tensor $\psi^{ik}$. After  four  gauge
conditions,  Eq.14 describes a mixture of two particles which have
two sources
$$
\{\psi^{ik}\} = \{2\} \oplus \{0\} \Leftrightarrow
\{T^{ik}\} = \{2\} \oplus \{0\}
\eqno(16)
$$
So that eq.(14) can be written in the form
$$
\Box\psi^{ik}_{\{2\}} =
                     \frac{8\pi G}{c^2}\,
                                  T^{ik}_{\{2\}}
                                           \quad
            \mbox{\rm or}
                                           \quad
\Box\phi^{ik} =
                     \frac{8\pi G}{c^2}
     \left[ T^{ik}_{(\Sigma)}
                     -\frac{1}{4}
 \eta^{ik} \, T_{(p)}\right]
\eqno(17a)
$$
and
$$
\Box\psi^{ik}_{\{0\}} =
                   - \frac{8\pi G}{c^2}\,
                                  T^{ik}_{\{0\}}
                                           \quad
            \mbox{\rm or}
                                           \quad
\frac{1}{4}
\Box\psi \eta^{ik} =
                    - \frac{8\pi G}{c^2}\,
           T_{(p)}
                       \frac{1}{4}
\eta^{ik}
\eqno(17b)
$$
where
$
\psi^{ik} = \psi^{ik}_{\{2\}} + \psi^{ik}_{\{0\}} =
\phi^{ik} + \frac{1}{2}\eta^{ik}\psi
$
(and
the  same representation for
$
T^{ik}_{(\Sigma)}
$), besides we take into account
that gravitons of both  kinds are massless particles, i.e. have
traceless~EMTs
($
T_{(\Sigma)} = T_{p}
$).

 The variation of action integral~(4)
with respect to particle coordinates gives the equation of motion in
the gravitational  field$^{4,\, 7}$
$$
A^{i}_{k}\frac{du^k}{ds} =
 -B^{i}_{kl}u^k u^l
\eqno(18)
$$
where
\begin{eqnarray}
   A_{k}^{i} &=&
            \left(1 -\frac{1}{c^2} \psi_{ln}
     u^l u^n \right) \eta^{i}_{k} -
 \frac{2}{c^2}\psi_{kn}u^n u^i + \frac{2}{c^2}\psi^{i}_{k},
                                          \numb(19)\\
  B^{i}_{kl} &=& \frac{2}{c^2} \psi^{i}_{k,l} -
      \frac{1}{c^2} \psi^{\phantom{kl},i}_{k,l} -
  \frac{1}{c^2}\psi_{kl,n}u^n u^i  . \numb(20)
\end{eqnarray}
\eject

\noindent
{\it
   1.3. Static Weak Field}
\vskip 5mm

       In the case of spherically symmetric static~(SSS)   weak  field
the  first   approximation~EMT   has a   very   simple   form
$
T_{(\Sigma)}^{ik} =
{\rm diag}(\rho_{0} c^2,\,0,\,0,\,0)
$
and solution of Eq.14 is   Birkhoff's potential
$$
\psi^{ik} = \varphi_{_{\rm N}} \,{\rm diag} (1,\,1,\,1,\,1),
\eqno(21)
$$
where
$
\varphi_{_{\rm N}}
$
is Newtonian potential
($
\varphi_{_{\rm N}} = -GM/r
$
outside
the
gravitating
body).

The gravitational field~(21) can  be expressed as
the sum of the spin~2 and spin~0 parts
$$
\psi^{ik} = \psi^{ik}_{\{2\}} +  \psi^{ik}_{\{0\}} =
 \varphi_{_{\rm N}} \,
 {\rm diag}
(\myfrac{3}{2}, \,\myfrac{1}{2},\, \myfrac{1}{2},\, \myfrac{1}{2}) +
(-  \varphi_{_{\rm N}}) \,
 {\rm diag}
(\myfrac{1}{2},\, -\myfrac{1}{2},\, -\myfrac{1}{2},\, -\myfrac{1}{2}).
\eqno(22)
$$

       Using  the  SSS  solution~(21)  and    corresponding~EMT
expressions~(9),~(10),~(11) we find the  total  energy  densities  of
the system from~Eq.8
$$
T_{(\Sigma)}^{00} = T_{(p)}^{00} + T_{(int)}^{00} +
   T_{(g)}^{00} = \left(\rho_{0} c^2 + e \right) +
\rho_{0}\varphi_{_{\rm N}} + \frac{1}{8\pi G}
\left(\nabla \varphi_{_{\rm N}}\right)^2
\eqno(23)
$$
where
($\rho_{0}c^2 + e$)
is  the  rest  mass  and   the  kinetic  energy
densities,
$
\rho_{0}\varphi_{_{\rm N}}
$
is the interaction energy density,
$
\nabla \varphi_{_{\rm N}}^2/8\pi G
$
is  the energy
density of the gravitational field. The total  energy  of  the system
will be
$$
E_{(\Sigma)} =
\int T_{(\Sigma)}^{00} dV = E_0 + E_k + E_p
\eqno(24)
$$
where
$
E_{0} = \int(\rho_{0}c^2)\,dV
$
is the rest-mass energy,
$
E_{k} = \int(e)\,dV
$
is
the kinetic energy, and $E_{p}$
is the classical potential energy
that equals the sum of the interaction and the gravitational field
      energy:
$$
E_p = E_{(int)} + E_{(g)} =
 \int(\rho_0 \varphi_{_{\rm N}} +
 \frac{1}{8\pi G} (\nabla \varphi_{_{\rm N}})^2)\,
 dV = \frac{1}{2} \int \rho_0 \varphi_{_{\rm N}}\, dV
\eqno(25)
$$

       It is important to note that the canonical EMT (9) include both
EMTs for spin~2 and spin~0 gravitons. Hence the $T^{00}_{(g)}$
given by~(9) is the sum of energy densities of both components (see
for details$^6$).

     For external field of SSS body in~PN  approximation  the  only
needed correction is to take into account the energy  density  of  the
gravitational field. Hence  we  have  exactly  Eq.1b  and  nonlinear
addition for Birkhoff's
$\psi^{00}$ component is
$$
\psi^{00} =  \varphi_{_{\rm N}} + \frac{1}{2}
\frac{ (\varphi_{_{\rm N}})^2}{c^2}.
\eqno(26)
$$

       Substituting Birkhoff's potential (21)  into  (18)  and  taking
into account nonlinear~PN correction~(26) one gets the three
dimensional equation of motion for test particle
$$
\frac{d {\bf
v}}{dt} = - \Bigl(1 + \frac{v^2}{c^2} +
        4\frac{\varphi_{_{N}}}{c^2} \Bigr) \nabla\varphi_{_{N}} + 4
                 \frac{{\bf v}}{c} \Bigl(\frac{{\bf v}}{c} \cdot
      \nabla\varphi_{_{N}}\Bigr).  \eqno(27)
$$
For Eq.27 the
pericenter shift of test particle is
$$
\delta\phi = \frac{6\pi G
M_0}{c^2 a(1 - e^2)}, \eqno(28)
$$
in which the nonlinear
contribution~(26) provides 16.7\% of  the value~(28). Therefore
in~TFT there is a direct testing of the positive  energy density of
the gravitational field.

It  is  easy to  show$^{5,\, 6,\, 7}$  that  all  PN
       classical   relativistic gravitational effects have the  same
values  as  in  GR  but the other interpretations. For  example
substituting
Eq.22
in Eq.18 one finds that
usual Newtonian force is the sum  of  the attracting force (spin~2)
and repulsing force (spin~0)
$$
F = F_{\{2\}} + F_{\{0\}} =
-\frac{3}{2}m_{0} \nabla\varphi_{_{N}} +
\frac{1}{2}m_{0}\nabla\varphi_{_{N}} =
- m_{0}\nabla\varphi_{_{N}} = F_{_{N}}
\eqno(29)
$$
\vskip 10mm

\noindent
{\it
   1.4. Free Field}
\vskip 5mm

       Within the scope of TFT not  only  static,  but  also  variable
gravitational field contains the sum of spin~2 and spin~0  gravitons.
The field equations in the form~(17), can  be
written
for the free field  as
$$
\Box\phi^{ik} = 0,  \quad
\phi^{ik}_{\phantom{ik},k} = 0
\eqno(29a)
$$
and
$$
 \frac{1}{4} \Box \psi \eta^{ik} = 0 .
\eqno(29b)
$$

Eqs.29 can be derived from the free field Lagrangians
$$
\Lambda_{\{2\}} =
             \frac{1}{16\pi G}
\phi_{lm,n}\phi^{lm,n} ,
\eqno(30a)
$$
and
$$
\Lambda_{\{0\}} =
             \frac{1}{64\pi G}
\psi_{,n}\psi^{,n}       .
\eqno(30b)
$$
Eqs.30 is a consequence of the gravitational  field  Lagrangian~(5)
for the field connected with sources (full field equations~(13)).  Note
that sign  of~(30b)  is  positive  due  to  positive  energy  density
condition for spin~0 free particles.  Corresponding~EMTs  for  tensor
(spin~2) and scalar (spin~0) gravitational waves are
$$
T_{\{2\}}^{ik} =
             \frac{1}{8\pi G}
\phi_{lm}^{\phantom{lm},i}\phi^{lm,k} ,
\eqno(31a)
$$
and
$$
T_{\{0\}}^{ik} =
             \frac{1}{32\pi G}
\psi^{,i}\phi^{,k}  .
\eqno(31b)
$$
\vskip 10mm
\noindent
{\bf
2. Emission of Tensor and Scalar Gravitational Waves}
\vskip 4mm

       Let us consider the retarded potentials solution of Eqs.17
in the wave zone for slow motions in the source.
For Eq.17a we  have usual quadrupole radiation (spin~2 gravitons)
with luminosity
$$ L_{\{2\}} = \frac{G}{45c^5}
{\stackrel{\ldots}{D}}_{\alpha\beta}^{
2}
\eqno(32)
$$
where
$
D_{\alpha\beta}
$
is the reduced quadrupole moment of the system of
       particles.

       In the same conditions the solution of the
       Eq.17b is
$$
\psi ({\bf r}, \, t) \approx
                         \frac{2GM_{0}}{r}
                        - \frac{2GE_{k}}{c^2r} +
                           \frac{2GM_{0}}{cr}
      \left({\bf n} \cdot \dot {\bf R} \right)
                            + \frac{G}{c^2r}
n_{\alpha}n_{\beta} \ddot{I}_{\alpha \beta}  ,
\eqno(33)
$$
where
$
M_{0} =\sum m_{a}
$  is the rest mass,
$
E_{k} = 1/2 \sum m_{a}v_{a}^{2}
$  is the
kinetic energy,
$
{\bf R} = \sum m_{a}{\bf r}_{a} / \sum m_{a}
$  is
the center of mass,
$
I_{\alpha\beta} = \sum m_{a} x_{a}^{\alpha} x_{a}^{\beta}
$
is the  moment of
inertia of the system.
Differentiation with respect to
the time  of Eq.33 leaves the only monopole term
$$
\dot{\psi}
({\bf r}, \, t) \approx - \frac{2G\dot{E}_{k}}{c^2r}
\eqno(34)
$$
because the first term equals to zero  in  consequence
of  rest mass conservation, the third term equals to zero in view of
inertial motion of the center of mass and the fourth term vanishes so
that it has been taken into account within  quadrupole
radiation~(32).  Substituting
Eq.34 into scalar wave~EMT~(31b) and
averaging it on the sphere  one gets the scalar monopole
luminosity (spin~0 gravitons)
$$
L_{\{0\}} = \frac{G}{2c^5} \dot
E_{k}^{2} \eqno(35)
$$
\vskip 8mm
\noindent
{\bf
3. Detection of Tensor
and Scalar Gravitational Waves
}
\vskip 4mm

\noindent
{\it
   3.1. Test Particles in Tensor Wave}
\vskip 4mm

       Let us  consider  inertial  frame  of  reference  where  we
analyzed action integral (4) and derived equations of motion for  test
particles~(18).  We  can   fix   Cartesian   coordinates   with
being the gravitational radiation source in the  origin  and  the
$x$-axis  along propagation  of  gravitational  wave.  For  plane
tensor   wave
$
\phi^{ik}
$
satisfying Eq.29a the only nonzero
components are
$
\phi^{22} = - \phi^{33} = A_{+}
$
 and
$
\phi^{23} =
 \phi^{32} = A_{\times}
$,
where for monochromatic wave
$
A(t,\,x) =
A^{0} \cos(\omega t  - kx)
$.  Substituting above expressions  into
Eq.18  and  leaving the  main terms  one  gets equation  of motion
of test particle in the plane monochromatic tensor wave
$$
\frac{dv_{x}}{dt} = 0   \quad
\mbox{ and }              \quad
\left\{
\begin{array}{c}
\displaystyle \frac{dv_{y}}{dt} =
                       2v_{y}
\frac{\dot{A}_{+}}{c^2} +
                       2v_{z}
\frac{\dot{A}_{\times}}{c^2}
                                \\ \\
             \displaystyle \frac{dv_{z}}{dt} =
                       2v_{y}
\frac{\dot{A}_{\times}}{c^2} -
                       2v_{z}
\frac{\dot{A}_{+}}{c^2}
\end{array}
\right. .
\eqno(36)
$$
It follows from Eq.36
that the tensor wave  is  transversal,~i.e. there is
no
acceleration along propagation  direction.  Besides  for the rest
test particle
($
v_{x} = v_{y} = v_{z} = 0
$) there is no
interaction between particle and tensor
wave
(in a manner like the
rest  charged particle  has no interaction with magnetic field in
electrodynamics).  Therefore  in the case of tensor gravitational wave
the velocity of test particle plays the role of ``gravitational
charge" and we need
such detectors in which  there is  system of particles
with nonzero relative velocities. For example  it may be rotating
bodies and even metallic bars where the  degenerate electrons
have nonzero  velocities relatively crystallic  proton lattice. The
amplitude of the velocity deviation can be  estimated  as
$
\Delta v =
2vA/c^2
$.
\vskip 9mm

\noindent
{\it
   3.2. Test Particles in Scalar Wave}
\vskip 4.5mm

       For  the  scalar   plane   monochromatic   gravitational   wave
satisfying Eq.29b we have
$
\psi^{ik}_{\{0\}} = A(t,\,x)\eta^{ik} = A^{0}
\cos(\omega t - kx)\eta^{ik}
$.
If we substitute this expression into Eq.18 and
leave the main terms we get the following equation of motion of test
particle in scalar wave
$$
\frac{dv_{x}}{dt} =
c\, \frac{\dot{A}}{c^2} \quad
\mbox{\rm and}    \quad
\left\{
\begin{array}{c}
\displaystyle \frac{dv_{y}}{dt} = 0  \\ \\
\displaystyle \frac{dv_{z}}{dt} = 0
\end{array}
\right. .
\eqno(37)
$$

According to Eq.37 the scalar wave is  longitudinal  and  the
rest test particle oscillates around initial  position
with the velocity
amplitude $\Delta v =  cA/c^{2}$   and
the distance  amplitude
$\Delta x  = A/kc^2$.  For two particles at a distance
$l_{0} \ll \lambda$ along
x-axis we get the dimensionless amplitude of oscillation in the form
$\Delta l_{0}/l_{0} = A/c^2$.  Hence the detection of the scalar
wave can be achieved by means of bar detectors and  laser
interferometric detectors.  It  is important  to  note  that  scalar
gravitational wave does not interact with  the  electromagnetic
field because the interaction Lagrangian is
$\psi^{ik}_{(0)} T_{ik} = \frac{1}{4} \psi\eta_{ik}T^{ik}_{(em)} = 0$.

\vskip 9mm
\noindent
{\bf
4. Binary Pulsar PSR 1913+16}
\vskip 4.5mm

\noindent
{\it
   4.1. Gravitational Radiation from Binary System}
\vskip 4.5mm

       For a  binary  system  the  loss  of  energy  due  to  tensor
gravitational radiation is given by Eq.32. Direct calculations  lead
to the well-known formula for  quadrupole  luminosity  of  the  binary
system
$$
<\dot{E}>_{\{2\}} = \frac{
                     32 G^{4} m_{1}^{2} m_{2}^{2}
    \left(m_1 + m_2 \right)
    \left(1 + \frac{73}{24}e^2 +
\frac{37}{96}e^4
                     \right)}
                    {5c^5 a^5
    \left(1 -e^2
                     \right)^{7/2}}    ,
\eqno(38)
$$
where $m_{1}$, $m_{2}$  are masses of the two bodies, $a$ is the
semimajor axis and $e$ is eccentricity of the relative orbit.
The similar
calculations based on Eq.35 give the loss  of  energy due to
scalar monopole radiation$^8$
$$
<\dot{E}>_{\{0\}} =   \frac{G^4 m_1^2 m_2^2
    \left(m_1 + m_2 \right) \left(e^2 +
        \frac{1}{4}e^4 \right) }
    {4c^5 a^5 \left(1 -e^2 \right)^{7/2}}   ,
\eqno(39)
$$
 The ratio of the scalar~(39) to the tensor~(38) luminosity is
$$
  \frac{<\dot{E}>_{\{0\}}}
         {<\dot{E}>_{\{2\}}} =
         \frac{5}{128} \cdot
         \frac{ \left(e^2 +
                           \frac{1}{4}e^4
           \right) }
                  { \left(1 + \frac{73}{24}e^2 +
                             \frac{37}{96}e^4
           \right) }.
\eqno(40)
$$
Note that the value
of the ratio~(40) lies in the interval $[0,\,1.1\%]$ and for the
circular orbit equals to zero.  However  for  a  pulsating
spherically symmetric body there is no quadrupole radiation and
scalar radiation becomes decisive.  In particular it follows from
this that in the~TFT it is impossible to have a ``quiet"
relativistic collapse of  a spherical body.
\vskip 10mm

\noindent
{\it
   4.2. PSR $1913+16$}
\vskip 5mm

       According to$^{9,\, 10}$ for the  binary  pulsar  PSR~$1913+16$
the eccentricity  $e = 0.6171309(6)$, hence the expected scalar
radiation part~(40) is 0.735~\%. As far as  the rate of orbital period
change $\dot{P}$ is proportional to the total energy loss
$ (<\dot{P}> = -\frac{3}{2}
\frac{E}{P}<\dot{E}>) $, one expectes the corresponding  excess of
orbital period decrease due to  scalar gravitational radiation.  The
observed value of the orbital period change is$^10$
$$
\dot{P}^{({\rm obs})} = - 2.425(10) \cdot 10^{-12},
$$
while the theoretical prediction for pure tensor
quadrupole radiation is$^9$
$$
\dot{P}^{({\rm theor})}_{\{2\}} = - 2.402576(69) \cdot 10^{-12}.
$$
Therefore the observed excess of the orbital period decrease is
$$
\Bigl({({\rm Obs})} - {({\rm Theor})}\Bigr) = + 0.96~\% \pm 0.4~\%
$$
that is very close to  the  expected value
0.735~\%  for  the  scalar gravitational radiation. Taking into
account the spin~0 part~(39) of the gravitational radiation we get
the theoretical prediction
$$
\dot{P}^{({\rm theor})} = \dot{P}^{({\rm theor})}_{\{2\}} +
\dot{P}^{({\rm theor})}_{\{0\}} = - 2.420254(69) \cdot 10^{-12}
$$
        in a good agreement with the observed value.
\vskip 10mm

\noindent
{\it
   4.3. Problem of Galactic Acceleration}
\vskip 5mm

       It has been shown in$^9$ that one must take  into  account  the
effect of the galactic acceleration of the pulsar  and  the  Sun,  and
that of the proper motion of the pulsar. Note that the  distance
$d$ to
PSR~$1913+16$ is the very sensitive parameter for the calculation of the
galactic effect. Unfortunately the line of sight to  the  pulsar  pass
through very complex region of our Galaxy and one must be very careful
to use known distances to other pulsars for distance  estimation  to
PSR~$1913+16$.  Indeed  in$^9$  it  was  used  indirect  arguments  to
re-estimate the standard dispersion-measure distance of 5.2~kpc and  to
get the new distance of~8.3 kpc. For the $d  =  8.3$  kpc  the
galactic effect is  $+0.69$~\%. However another
arguments$^{11}$ based on an
analysis  of the pulse structure of  PSR~1913+16 itself lead to an
estimation  of the distance about 3~kpc. For the $d = 3$~kpc the
galactic  effect is   $+0.11$~\%.  In  any  case  it  is  clear  that
we  need   further investigation of the pulsar and the direct
estimation of the  distance to PSR~$1913+16$.
\vskip 10mm

\noindent
{\bf
5. References}
\vskip 5mm

\leftskip 12mm
\parindent 0pt
\parbox{0pt}{\hidewidth 1.\hbox{\ }}C.Misner,
K.Thorn,  J.Wheeler,  {\it Gravitation}  (W.H.Freeman,  San
      Francisco, 1973), Chapters~7 and~18.

\parbox{0pt}{\hidewidth 2.\hbox{\ }}L.Landau,
E.Lifshiz, {\it Field Theory} (Nauka, Moscow, 1973).

\parbox{0pt}{\hidewidth 3.\hbox{\ }}N.Bogolubov,
D.Shirkov, {\it In\-tro\-duc\-tion  to  Quan\-tum  Field  Theo\-ry}
      (Nau\-ka, Mos\-cow, 1976).

\parbox{0pt}{\hidewidth 4.\hbox{\ }}G.Kalman,
{\it Phys.Rev.}, {\bf 123}, (1961), 384.

\parbox{0pt}{\hidewidth 5.\hbox{\ }}W.E.Thirring,
{\it Ann.Phys.}, {\bf 16}, (1961), 96.

\parbox{0pt}{\hidewidth 6.\hbox{\ }}V.Sokolov,
Yu.Baryshev, {\it Grav.and Rel.Theor.},
KGU, vyp.{\bf 17}, (1980), 34.

\parbox{0pt}{\hidewidth 7.\hbox{\ }}Yu.Baryshev,
{\it Vestnik LGU}, ser.1, vyp.{\bf 4}, (1986), 113.

\parbox{0pt}{\hidewidth 8.\hbox{\ }}Yu.Baryshev,
{\it Astrophysics}, {\bf 18}, (1982), 58.

\parbox{0pt}{\hidewidth 9.\hbox{\ }}T.Damour,
J.Taylor, {\it Astrophys.J.}, {\bf 366}, (1991), 501.

\parbox{0pt}{\hidewidth 10.\hbox{\ }}J.Taylor
et al., {\it Nature}, {\bf 355}, (1992), 132.

\parbox{0pt}{\hidewidth 11.\hbox{\ }}Yu.Baryshev,
A.Pynzar, (in preparation).
\end{document}